\documentclass[final,3p,times]{elsarticle}

 \usepackage{graphics,epsfig,subfigure}

\usepackage{amssymb}
\usepackage{cases}

\usepackage{amstext}
\usepackage{color}
\usepackage{array}
\usepackage{multirow}
\newcommand\beq{\begin{equation}}
\newcommand\eeq{\end{equation}}
\newcommand\beqn{\begin{eqnarray}}
\newcommand\eeqn{\end{eqnarray}}

\newcommand\fc{\frac}
\newcommand\lt{\left}
\newcommand\rt{\right}
\newcommand\pt{\partial}
 \biboptions{comma,square,compress,numbers,sort}

\journal{Physics Letters B}

\begin{document}

\begin{frontmatter}



\title{Domain wall brane in a reduced Born-Infeld-$f(T)$ theory}


\author[label1]{Ke Yang}
 \author[label2,label3]{Wen-Di Guo}
  \author[label2,label3]{Zi-Chao Lin}
  \author[label2,label3,label4]{Yu-Xiao Liu\corref{cor1}}
  \ead{liuyx@lzu.edu.cn}
  \cortext[cor1]{The corresponding author.}

\address[label1]{School of Physical Science and Technology, Southwest University, Chongqing 400715, China}
\address[label2]{Research Center of Gravitation, Lanzhou University, Lanzhou 730000, China}
\address[label3]{Institute of Theoretical Physics, Lanzhou University, Lanzhou 730000, China}
\address[label4]{Key Laboratory for Magnetism and Magnetic of the Ministry of Education, Lanzhou University, Lanzhou 730000, China}

\begin{abstract}

The Born-Infeld $f(T)$ theory is reduced from the Born-Infeld determinantal gravity in Weitzenb\"ock spacetime. We investigate a braneworld  scenario in this theory and obtain an analytic domain wall solution by utilizing the first-order formalism. The model is stable against the linear tensor perturbation. It is shown that the massless graviton is localized on the brane, but the continuous massive gravitons are non-localized and will generate a tiny correction with the behavior of ${1}/{(k r)^{3}}$ to the Newtonian potential. The four-dimensional  teleparallel gravity is recovered as an effective infrared theory on the brane. As a physical application, we consider the (quasi-)localization property of spin-1/2 Dirac fermion in this model. 
\end{abstract}


\end{frontmatter}



\section{Introduction}

In 1928, Einstein proposed a theory in order to unify the gravity and electromagnetism \cite{Einstein1928}, which is well known as the teleparallel equivalent of general relativity (TEGR or teleparallel gravity for short). It allows us to interpret the general relativity as a gauge theory for a translation group. Instead of a torsion-free pseudo-Riemannian spacetime, the underlying spacetime in this theory is a curvature-free Weitzenb\"ock  spacetime, which is characterized by the Weitzenb\"ock connection defined  in terms of a dynamical vielbein (or vierbein/tetrad in four-dimensional spacetime) field ${e^{A}}_{M}$.  The vielbein is an orthonormal basis in the tangent space of each spacetime point and relates to the spacetime metric field via $g_{MN}=\eta_{AB}{e^{A}}_{M}{e^{B}}_{N}$, with $\eta_{AB} = \text{diag}(-1, 1, 1, 1,1)$ the Minkowski metric for the tangent space. Here and after, the capital Latin indices $A,B,\cdots$ and $M,N,\cdots$ label the five-dimensional coordinates of tangent space and spacetime respectively, and the small Latin indices $a,b,\cdots$ and Greek indices $\mu,\nu,\cdots$ label the four-dimensional coordinates of tangent space and spacetime respectively.

The corresponding Weitzenb\"ock connection is given by
${\Gamma^{P}}_{MN}={e_{A}}^{P}\pt_{N}{e^{A}}_{M}$,
and the torsion of a spacetime is given by
${T^{P}}_{MN}={\Gamma^{P}}_{NM}-{\Gamma^{P}}_{MN}$.
The difference between the Weitzenb\"ock connection ${\Gamma^{P}}_{MN}$ and Levi-Civita connection ${\{^{P}}_{MN}\}$ is known as the contorsion tensor ${K^{P}}_{MN}\equiv {\Gamma^{P}}_{MN}-{\{^{P}}_{MN}\}
=\fc{1}{2}({{T_{M}}^{P}}_{N}+{{T_{N}}^{P}}_{M}-{{T^{P}}_{M}}_{N})$.
By using the torsion and contorsion tensors, one defines a new tensor ${S_P}^{MN}\equiv\fc{1}{2}(K^{MN}{}_{P}+\delta^{N}_{P}T_{Q}{}^{QM}
-\delta^{M}_{P}T_{Q}{}^{QN})$,
then a torsion scalar called Weitzenb\"ock invariant is constructed as
$T={S_P}^{MN}{T^{P}}_{MN}$.
So the well-known action of teleparallel gravity reads
\beq
S_{\text{tel}}=-\fc{1}{16\pi G_*}\int{}d^{D}x~eT,
\label{Action_teleparallel}
\eeq
where $e=|{e^A}_M|=\sqrt{-|g_{MN}|}$ and we will set the $D$-dimensional gravitational constant $G_*$ to be $4\pi G_*=1$ for later convenience.  With the identity $T=-R+2e^{-1}\partial_N(e{T_M}^{MN})$ between the Weitzenb\"ock invariant and Ricci scalar, the equivalence between teleparallel gravity and  general relativity is manifest. However, in order to investigate the difference between pure torsion space and pure curvature space,  following the spirit of $f(R)$ theory, teleparallel gravity was  generalized into an $f(T )$ type theory, which can explain the observed accelerating universe without introducing the dark energy \cite{Bengochea2009,Bamba2012}.

Another important generalization of teleparallel gravity follows the spirit of the well-known Born-Infeld electromagnetic theory, which can remove the divergence of the electron's self-energy in classical dynamics \cite{Born1934}. In Ref. \cite{Deser1998}, Deser and Gibbons proposed a Born-Infeld type gravitational theory aiming to regularize some singularities in general relativity. Since then the Born-Infeld type gravities have drawn much attention, see Ref. \cite{BeltranJimenez2017} for review and references therein.

One of the Born-Infeld type generalizations of gravity can be arranged as the form \cite{Deser1998,Vollick2004,Comelli2005,Banados2010,Chen2016,Ferraro2007,Ferraro2008,Ferraro2010,Fiorini2013,Fiorini2016a}
\beq
I_{\text{BIG}}=\frac{\lambda}{4}\int{}d^Dx\lt[\sqrt{-|g_{MN}+2\lambda^{-1} F_{MN}|}-\Delta\sqrt{-|g_{MN}|} \rt],
\eeq
where the rank-2 tensor $F_{MN}=F_{MN}(\psi,\pt\psi,\cdots)$ is a function of a certain field $\psi$ and its derivatives.
In low-energy limit, the action reduces to
\beq
I_{\text{BIG}}\approx\frac{1}{4}\int{}d^Dx\sqrt{-|g_{MN}|}\lt[\text{Tr}(F_{MN})+(1-\Delta)\lambda\rt].
\eeq

The simplest case is that $\text{Tr}(F_{MN})$ equals to Ricci scalar, then the low-energy theory is just the general relativity. By working in the pure metric formalism \cite{Deser1998}, i.e., $F_{MN}=R_{MN}(g)$, the theory leads to fourth order field equations with ghost-like instabilities.  However, by working in the Palatini formalism \cite{Vollick2004,Banados2010}, i.e., $F_{MN}=R_{(MN)}(\Gamma)$, there is no higher order derivative of the metric in field equations. One intensively studied Palatini Born-Infeld type theory is the so-called Eddington-inspired Born-Infeld (EiBI) gravity, which may regularize some gravitational singularities in classical dynamics \cite{Banados2010,Pani2011,Pani2012,Olmo2013}.

However, there is another possibility that one can obtain a proper low-energy theory, i.e., $\text{Tr}(F_{MN})=-T$. Thus, the teleparallel gravity (so general relativity equivalently) is recovered in this case \cite{Ferraro2007,Ferraro2008,Ferraro2010,Fiorini2013}. An interesting Born-Infeld type  generalization of teleparallel gravity has been proposed by Fiorini  in the work \cite{Fiorini2013}, where $\Delta=1$ in order to vanish the cosmological constant and $F_{MN}=\alpha F^{(1)}_{MN}+\beta F^{(2)}_{MN}+\gamma F^{(3)}_{MN}$ with each $F^{(i)}_{MN}$ defined by $F^{(1)}_{MN}={S_M}^{PQ}T_{NPQ}$, $F^{(2)}_{MN}={S_{PM}}^{Q}{T^{P}}_{NQ}$, and $F^{(3)}_{MN}=g_{MN}T$. In order to recover the teleparallel gravity in low-energy limit, these constants have to satisfy $\alpha+\beta+D\gamma=-1$. So the gravitational action in $D$-dimensional Weitzenb\"ock spacetime is given by
\beq
I_{\text{BIG}}=\frac{\lambda}{4}\int{}d^Dx\lt[\sqrt{-|g_{MN}+2\lambda^{-1} F_{MN}|}-\sqrt{-|g_{MN}|} \rt].
\label{BIG_Action}
\eeq
A novel property of this theory is that it exhibits some regular solutions in early universe for certain region of the parameter space \cite{Fiorini2013,Fiorini2016}. A general analysis of cosmological singularities in other regions of parameter space was presented in Ref. \cite{Bouhmadi-Lopez2014a}. Moreover, Schwarzschild geometry was considered under this framework in Ref. \cite{Fiorini2016a}.

Especially, a simple but nontrivial case is that $\alpha=\beta=0$, then the action reduces to an $f(T)$ type theory 
\beq
I_{\text{BIG}}=\frac{\lambda}{4}\int{}d^Dx ~ e \lt[\lt(1-\frac{2T}{D\lambda}\rt) ^{D/2}-1\rt].
\label{Action_BI0}
\eeq
For the signature we used, by redefining the parameter $\lambda\rightarrow-\lambda$, the action is rewritten into a more conventional form \cite{Fiorini2013}
\beq
I_{\text{BI0}}=-\int{}d^Dx ~ ef(T)=-\frac{\lambda}{4}\int{}d^Dx ~ e \lt[\lt(1+\frac{2T}{D\lambda}\rt) ^{D/2}-1\rt],
\label{Action_BI0}
\eeq
A cosmological  application of this reduced Born-Infeld-$f(T)$ theory was considered in Ref. \cite{Jana2014}.

On the other hand,  the idea that our spacetime dimensions are possibly more than four in ultra-violate regime is a long historical topic since the proposal of Kaluza-Klein (KK) theory in 1920s. One of the scenarios is that our Universe is a 3-brane embedded into a higher-dimensional bulk \cite{Rubakov1983}. The braneworld scenario opens a new way to solve some long standing problems in particle physics and cosmology \cite{Arkani-Hamed1998,Arkani-Hamed2000,Randall1999,Randall1999a,Dvali1999,Langlois2006}. Besides the general relativity, the braneworld scenario has been generalized into some alternative gravities, see Ref. \cite{Liu2017} for introduction and references therein. 
Most interesting thick brane configurations are the topological defects embedded in a higher-dimensional curved spacetime. And one of the commonest defects is the domain wall, which is considered as a kind of smooth generalization of Randall-Sundrum-2 (RS2) model \cite{Randall1999a}. The domain wall is a higher-dimensional generalization of one-dimensional soliton, called as kink, which is a non-trivial map from the space boundaries to the set of scalar vacua  \cite{Rubakov1983}.   As is known, the domain wall brane has some nice features, such as 1) the massless graviton can be localized on the brane and the four-dimensional effective gravity can be recovered \cite{DeWolfe2000,Csaki2000}, 2) the four-dimensional massless chiral fermion is obtained by introducing a Yukawa coupling between the five-dimensional fermion and the background kink scalar \cite{Liu2017}, 3) the four-dimensional gauge boson preserving charge universality is possibly obtained in the domain wall via the Dvali-Shifman mechanism \cite{Dvali1997}, 4) the Higgs doublet can be localized on the brane through its potential coupling to the kink scalar \cite{Davies2008}. Therefore, the domain wall brane scenario could be a potential candidate and the Standard Model may possibly be transplanted into a domain wall brane \cite{Davies2008}.

The KK and RS2 model were reconstructed in teleparallel gravity in Refs. \cite{Geng2014} and \cite{Nozari2013}, respectively. In Ref. \cite{Fiorini2014}, cosmological scenarios were discussed  in $f(T )$ gravities for the toroidal and spherical compactified extra dimensions. The inflation and dark energy dominated stage were realized in four-dimensional effective $f(T)$ gravities originated from five-dimensional  KK and RS2 theories \cite{Bamba2013}. Moreover, in Ref. \cite{Yang2012b}, the authors investigated a domain wall  brane in $f(T)$ theory with $f(T)=T+k T^n$, where the brane will split when spacetime torsion goes strong. The author in Ref. \cite{Menezes2014} generalized the domain wall model with a non-standard scalar kinetic term, and obtained an analytical solution by utilizing the first-order formalism. The tensor perturbation of these $f(T)$ domain wall branes was presented in a recent work \cite{Guo2016}, where the authors proved that the brane systems are stable and that the massless graviton is localized on the brane.

In this work, we are interested in generalizing the domain wall braneworld scenario into the reduced Born-Infeld-$f(T)$ gravity. Resorting to the first-order formalism, we can solve the field equations analytically. The paper is organized as follows. In section \ref{Model}, we build up a five-dimensional braneworld model and solve the system. In section \ref{Perturbation}, we analyze stability of the model under the linear tensor perturbation, and study the localization of KK gravitons and  Newtonian potential on the brane. In section \ref{Fermion_Resonances}, we investigate the (quasi-)localization property of spin-1/2 Dirac fermion based on this model. Finally, brief conclusions are presented.

\section{Domain wall brane solution} \label{Model}

In order to investigate the braneworld solution in this framework, we fix the background spacetime to be five dimensions ($D=5$) and consider the brane configuration built by a background scalar field, i.e., $\mathcal{L}_{\text{M}}=-\frac{1}{2}\pt^M\phi\pt_M\phi-V(\phi)$, which couples to the gravity in a conventional way. So the full action reads
\beq
I_{\text{tot}}=I_{\text{BI0}}+\int{}d^5x~e\mathcal{L}_{\text{M}},
\label{Action_tot}
\eeq
where $I_{BI0}$ is given by Eq. (\ref{Action_BI0}).
By varying this action with respect to the vielbein and the scalar  respectively, one obtains the field equations \cite{Yang2012b},
\begin{subequations}\label{EOM}
\beqn
e^{-1}f_T g_{NP}\pt_Q \lt(e{S_M}^{PQ}\rt)+f_{TT}{S_{MN}}^{Q}\pt_Q T-f_T{{\Gamma}^P}_{QM}{S_{PN}}^Q+\frac{1}{4}g_{MN}f(T) &=& \mathcal{T}_{MN},\label{EOM1}\\
{e}^{-1}\pt^K\lt(e \pt_K \phi \rt) &=& \frac{\pt V}{\pt\phi},\label{EOM2}
\eeqn
\end{subequations}
where the notations $f_T\equiv\pt f/\pt T$, $f_{TT}\equiv\pt^2 f/\pt T^2$, and $\mathcal{T}_{MN}$ is the energy-momentum tensor of the matter field.

The most general metric ansatz keeping the four-dimensional Poincar\'e invariance is given by
$ds^2=e^{2A}\eta_{\mu\nu}dx^\mu dx^\nu+dy^2$,
where $e^{A(y)}$ is the so-called warp factor.  Then the corresponding proper vielbein reads ${e^A}_M=\text{diag}\lt(e^A,e^A,e^A,e^A,1 \rt)$. The four-dimensional Poincar\'e symmetry constrains the background scalar field to depend on the extra dimension $y$ only.
With the metric and vielbein ansatz, the field equations (\ref{EOM}) read explicitly as
\begin{subequations}\label{EOM_Exp}
\beqn
6A'^2 f_T+\frac{1}{4}f &=& \frac{1}{2}\phi'^2-V,\label{EOM_Exp1}\\
36A'^2A''f_{TT} -(\frac{3}{2}A''+6A'^2)f_T-\frac{1}{4}f &=& \frac{1}{2}\phi'^2+V,\label{EOM_Exp2}\\
\phi''+4A'\phi'  &=& \frac{dV}{d\phi},\label{EOM_Exp3}
\eeqn
\end{subequations}
where the prime denotes the derivative with respect to the extra dimension $y$.
There are three unknown variables $A$, $\phi$, and $V(\phi)$, but only two of the three equations of motion are independent, so the system is underdetermined. Because of the non-linear expression of $f(T)$, it is hard to solve the theory directly by inputing some constraint. So we resort to the first-order formalism, which transforms the equations of motion into first-order equations via introducing a superpotential.

The first-order derivative of the warp factor is assumed to be  a function of $\phi$, called superpotential $W(\phi)$, i.e., $A'(y)=-\frac{W(\phi)}{3}$.
Then from the first two equations in (\ref{EOM_Exp}), we have the following relations
\beqn
\phi'(y)&=&\frac{ {X^{\frac{1}{2}} {{(\phi )}}Y(\phi )W'(\phi )}} {{30\sqrt {15} {\lambda ^{3/2}}}},\label{Exp_phiprime}\\
V (\phi)&=&\frac{\lambda }{4}-\frac{X^{3/2}(\phi )Z(\phi )}{900\sqrt {15} \lambda ^{3/2}} + \frac{X(\phi )Y^2(\phi ) W'^2{{(\phi )}}}{{27000{\lambda ^3}}} , \label{Exp_V}
\eeqn
where $X(\phi )=15\lambda  - 8W^2{{(\phi )}} $, $Y(\phi )=15\lambda  - 32W^2{{(\phi )}}$, and  $Z(\phi )=15\lambda  + 32W^2{{(\phi )}}$. Now with a given superpotential, the full theory can be solved from these first-order equations.

Because we are interested in thick brane with topological defect of the domain wall, the scalar field $\phi(y)$ is a kink soliton, which is an odd function of extra dimension. Besides, the bulk should be $Z_2$ symmetric along the extra dimension in order to recover the massless chiral fermions on the brane. To be consistent with the $Z_2$ symmetric bulk,  the warp factor $A(y)$ must be an even function of extra dimension. Therefore, to fulfill these conditions, a given superpotential $W(\phi )$ should be an odd function of the scalar $\phi$.

We can get an analytic brane solution from the superpotential ansatz
$W(\phi)=\frac{2 k  \phi}{\sqrt{3}}$,
 where a mass dimension one parameter $k$ has been introduced by $\lambda\equiv\frac{32}{45}k^2$ in order to simplify expressions and to count the dimension. The solution reads
 \begin{subequations}\label{Brane_Solution}
 \beqn
 \phi(y)&=&\frac{\tanh\lt(ky\rt)}{\sqrt {3+\tanh^2\lt(ky\rt) }},\\
 A(y)&=&-\frac{1}{3 \sqrt{3}}{\ln \left[\left(2-\sqrt{3}\right)  \left(2+\sqrt{4-\text{sech}^2(k y)}\right)\cosh (k y)\right]},\\
 V(\phi)&=&\frac{ k^2}{90} \left[31-16 \sqrt{1-\phi ^2}-\left(3-4 \phi ^2\right) \left(45-60 \phi ^2+16 \sqrt{1-\phi ^2}\right)\phi ^2 \right].
 \eeqn
 \end{subequations}
It shows that when the extra dimension $y$ runs from one boundary $y\rightarrow-\infty$ to the other $y\rightarrow\infty$, the scalar field $\phi(y)$ runs smoothly from $\phi(-\infty)\rightarrow-\frac12$ to $\phi(\infty)\rightarrow\frac12$, where the potential vacua  $V_0(\pm\frac12)=\frac{2(4-3\sqrt{3})}{45}k^2$ are just located. So the scalar is indeed a kink solution. 

\section{Tensor perturbation and gravitons}\label{Perturbation}

In this section, we  investigate the property of four-dimensional KK gravitons in this domain wall braneworld. 
In the standard scenario of metric perturbations, $ds^2=(g_{MN}+h_{MN})dx^M dx^N$, the perturbed part can be decomposed into $h_{55}=2\psi, h_{\mu5}=e^{A}(\pt_\mu F+G_\mu), h_{\mu\nu}=e^{2A}[2\eta_{\mu\nu}\varphi+2\gamma_{\mu\nu}+2\pt_\mu\pt_\nu B+2\pt_{(\mu} C_{\nu)}]$ with $\psi$, $\varphi$, $F$, and $B$ the scalar modes, $G_\mu$ and $C_\nu$ the transverse vector modes, and $\gamma_{\mu\nu}$ the transverse-traceless (TT)  tensor mode. There are totally 15 degrees of freedom, but the diffeomorphism of $x^M\rightarrow x^M+\epsilon^M(x^N)$ indicates further that there are gauge freedoms in the scalar and vector modes while the tensor mode is gauge independent. 

However, since the local Lorentz invariance is broken in $f(T)$ theory,  the broken gauge freedom in tangent frame will release some extra degrees of freedom in the vielbein \cite{Li2011d}. More precisely, there are 6 extra degrees of freedom for the four-dimensional case and 10 for the five-dimensional.  This can be seen easily from the evidence that the field equation (\ref{EOM1}) can be rewritten as $f_T G_{MN}+\frac12 g_{MN}(f-Tf_T)+f_{TT} S_{NMK}\nabla^K T=2\mathcal{T}_{MN}$ with $G_{MN}$ the Einstein tensor, and there is a constraint equation from the antisymmetry part of the field equation, i.e., $ (S_{NMK}-S_{MNK})f_{TT} \nabla^K T=0$. The constraint equation vanishes identically for the background vielbein ansatz ${e^A}_M=\text{diag}\lt(e^A,e^A,e^A,e^A,1 \rt)$ just like the FRW geometry. But  this is not the case for the vielbein perturbations. The constraint equation governs the dynamics of the extra degrees of freedom in perturbations. Thus, one can generally write the perturbed vielbein as
${e^A}_{M}={\bar e}^A{}_{M}+{\ae^A}_{M}$, where ${\bar e}^A{}_{M}$ corresponds to the degrees of freedom of the metric  $g_{MN}$, which satisfies the condition $g_{MN}=\eta_{AB} {e^A}_{M} {e^A}_{N}=\eta_{AB}{\bar e}^A{}_{M}{\bar e}^A{}_{N}$, while ${\ae^A}_{M}$ illustrates the 10 extra degrees of freedom. Up to the linear order, the perturbed vielbein can be decomposed into \cite{Wu2012a,Izumi2013}:
${\bar e^5}_{5}=1+\psi$, ${\bar e^a}_{5}=0$, 
${\bar e^5}_{\mu}=e^{A}(\pt_\mu F+G_\mu)$,  
${\bar e^a}_{\mu}=e^{A}\delta^{a\nu}[(1+\varphi)\delta_{\mu\nu}+\pt_\mu\pt_\nu B+\pt_{(\mu} C_{\nu)}+\gamma_{\mu\nu} ]$, and ${\ae^5}_{5}=0$, ${\ae^a}_{5}=\delta^{a\mu}(\pt_\mu \beta+D_\mu)$, ${\ae^5}_{\mu}=0$, ${\ae^a}_{\mu}=\delta^{a\nu}\epsilon_{\mu\nu\lambda}(\pt^\lambda \sigma+V^\lambda)$, with $\beta$ a scalar, $D_\mu$ a transverse vector, $\sigma$ a pseudoscalar, and $V^\lambda$ a transverse pseudovector. Thus, the 10 extra degrees of freedom are only encoded in scalar and transverse vector modes. The $f(T)$ theory is also invariant under the coordinate transformation $x^M\rightarrow x^M+\epsilon^M(x^N)$, and this provides gauge choices to eliminate some of the scalar and transverse vector modes. After scalar-vector-tensor decomposition, the scalar, transverse vector, and TT tensor modes can be decoupled from each other, and one can deal with them separately.

Since the spin-2 gravitons associate with the TT tensor perturbation, we close all the scalar and transverse vector modes, and only focus on the TT tensor mode in perturbed metric, i.e., $ds^2=e^{2A}(\eta_{\mu\nu}+2\gamma_{\mu\nu})dx^\mu dx^\nu+dy^2$. Correspondingly, the perturbed vielbein involving only the TT tensor mode reads
\beq
{e^A}_M=\left( {\begin{array}{*{20}{c}}
e^{A(y)}\delta^{a\nu}\left(\delta_{\mu\nu}+{\gamma_{\mu\nu}} \right)&  0\\
0  & 1
\end{array}} \right),
\eeq
where the TT tensor perturbation $\gamma_{\mu\nu}=\gamma_{\mu\nu}(x^\mu,y)$ satisfies the TT condition $\pt^\mu\gamma_{\mu\nu}=\eta^{\mu\nu}\gamma_{\mu\nu}=0$. Now the constraint equation vanishes automatically for this vielbein choice. By substituting the perturbed vielbein into the field equations (\ref{EOM}), and making some cumbersome but simple algebra, one gets an equation governing the TT tensor perturbation \cite{Guo2016}
\beq
\gamma_{\mu\nu}''+4A'\lt(1-6A''\frac{f_{TT}}{f_T}\rt)\gamma_{\mu\nu}'+e^{-2A}\Box^{(4)}\gamma_{\mu\nu}=0,
\label{TTEquation_y}
\eeq
where $\Box^{(4)}=\eta^{\mu\nu}\pt_\mu\pt_\nu$ is the four-dimensional d'Alembert operator. Further, with a coordinate transformation $dz=e^{-A}dy$, the factor $e^{-2A}$ before the d'Alembert operator is eliminated and the equation reads
\beq
\lt(\pt_z^2+2K(z)\pt_z+\Box^{(4)} \rt)\gamma_{\mu\nu}=0,
\label{TTEquation_z}
\eeq
where 
$K(z)\equiv\frac{3}{2}\pt_z A+12e^{-2A}\lt((\pt_z A)^2-\pt_z^2 A\rt)\pt_z A{f_{TT}}/{f_T}$.
Finally, by imposing a KK decomposition $\gamma_{\mu\nu}(x^\rho,z) = \epsilon_{\mu\nu}(x^\rho)\exp\lt(-\int K(z) dz\rt)\Psi(z)$,  Eq.~(\ref{TTEquation_z}) reduces to a four-dimensional Klein-Gordon equation $(\Box^{(4)}-m^2)\epsilon_{\mu\nu}=0$ owing to the preserved four-dimensional Poincar\'e symmetry, and an aimed Schr\"odinger-like equation
\beq
\lt(-\pt_z^2+U(z)\rt)\Psi=m^2\Psi,
\eeq
where the effective potential reads $U(z)=K^2+\pt_z K$. The Hamiltonian in the above equation can be factorized into a form of supersymmetric quantum mechanics, i.e.,
$H(z)=\mathcal{K}^\dagger\mathcal{K}=\lt(\pt_z+K\rt)\lt(-\pt_z+K\rt)$.
Given a Neumann boundary condition $\pt_z\gamma_{\mu\nu}|_\text{Boundary}=0$, all the eigenvalues $m^2$ are easily proved to be non-negative \cite{Yang2017}. So there is no tachyonic graviton and hence the model is stable under the linear TT tensor perturbation.
The mass spectrum of KK gravitons is totally determined by the effective potential, and there is a manifest massless graviton given by
\beq
\Psi_{0}=\exp\lt(\int{ K dz}\rt)=\lt(1-\frac{27}{4 k^2}A'(y)^2\rt) ^{\frac{3}{4}} e^{\frac{3}{2}A(y)},
\eeq
 up to a normalization coefficient $N_0$. It is clear that the massless graviton only propagates on the brane, i.e., $\gamma_{\mu\nu}^{0}(x^\rho)=N_0\epsilon_{\mu\nu}^{0}(x^\rho)$.  So the massless graviton contributes to the four-dimensional gravity.
Then by including only the massless mode in the metric $ds^2=e^{2A}\tilde g_{\mu\nu}(x)dx^\mu dx^\nu+dy^2$ with $\tilde g_{\mu\nu}(x)=\eta_{\mu\nu}+\gamma^{0}_{\mu\nu}(x)$, the five-dimensional Weitzenb\"ock invariant can be rewritten as $T=e^{-2A}\tilde T-12A'^2$, where the four-dimensional invariant $\tilde T$ is calculated from $\tilde g_{\mu\nu}(x)$.  Thus, the low-energy effective gravity on the brane can be read off from the action (\ref{Action_tot}),
\beq
I_{\text{tot}}\supset I_{\text{eff}}=-\frac{1}{16\pi G_N}\int d^{4} x~ \tilde{e} \tilde{T}+\cdots,
\eeq
where the effective four-dimensional gravitational constant $G_N$ is given by $G_N^{-1}=G_*^{-1} \int {dy\lt(1-\frac{27}{4 k^2}A'(y)^2\rt) ^{\frac{3}{2}} e^{2A(y)}}=G_*^{-1} \int {dye^{-A}\Psi_{0}^2}=G_*^{-1} \int {dz\Psi_{0}^2}\approx \frac{4.59}{k}G_*^{-1}$. It is clear that a normalizable massless mode ensures that the effective four-dimensional gravity can be recovered on the brane, and the normalization coefficient  reads $N_0\approx\sqrt{\frac{k}{4.59}}$. Therefore, the infrared gravitational theory on the brane is just the teleparallel gravity. By setting the parameter $k$ to be of order the Planck scale $ M_{\text{Pl}}$, the fundamental gravitational constant $G_*^{-1}\sim M_{\text{Pl}}^3$.  

The Schr\"odinger potential is a volcano-like one, which is universal for flat braneworlds. Therefore, besides one localized massless mode, there are continuous  massive KK gravitons starting from $m>0$ and all these modes can propagate along extra dimension. They will contribute to a correction to the Newtonian potential at short distance. However, if the massless ground state  is normalizable,  the continuous modes are naturally decoupled and do not lead to unacceptably large corrections to the Newtonian potential \cite{Csaki2000}. Quantitatively, for two point-like sources separated by a distance $r$ on the brane (say $z=0$ for simplicity), the correction is determined by asymptotic behavior of  the Schr\"odinger potential in large $z$ region \cite{Csaki2000,Bazeia2009}. If a volcano-like potential behaves like $U(z)\sim\alpha(\alpha+1)/z^2$ for $z\gg 1$, then the properly normalized wave functions of massive KK modes will take the form $|\Psi_{n}(0)|\sim \lt(\frac{m}{k}\rt)^{\alpha-1}$ on the brane, and the exchange of these massive gravitons between the two sources will yield a correction proportional to ${1}/{(k r)^{2\alpha}}$ to the Newtonian potential.  In our case, the Schr\"odinger potential behaves like $U(z)\sim\frac{15}{4z^2}$ for $z\gg 1$. This behavior is the same as that of the RS2 model \cite{Randall1999a}, but is different from that of the EiBI domain wall \cite{Liu2012,Fu2014}, where the Schr\"odinger potential behaves like $U(z)\sim\frac{(3+4 n) (5+4 n)}{4 z^2} $ with $n$ a positive integer. Thus, the correction to the Newtonian potential reads ${1}/{(k r)^{3}}$ in this model, which is very tiny provided that the distance $r$ between the two sources is much larger than the Planck length.

\section{KK resonances of spin-1/2 fermions}\label{Fermion_Resonances}

In this section, we focus on the (quasi-)localization property of spin-1/2 Dirac field on the brane. In a five-dimensional spacetime, fermions are still described by a four-component Dirac spinors. Its Clifford algebra reads $\{\Gamma^M,\Gamma^N\}=2g^{MN}$, where the five-dimensional Dirac matrices $\Gamma^M={e_B}^M\Gamma^B$, and $\Gamma^B=(e^{-A(z)}\gamma^{\mu},e^{-A(z)}\gamma^5)$ with $\gamma^{\mu}$ and $\gamma^5$ the usual flat gamma matrices in the Dirac representation. So the action for a massless spin-1/2 Dirac field coupling to the background scalar field in a Yukawa-type interaction is given by 
\beq
S_{\Psi}=\int{}dx^5e\left(\bar\Psi\Gamma^M(\pt_M+\omega_M)\Psi-\eta\phi\bar\Psi\Psi  \right),
\eeq
where $\eta$ is the Yukawa coupling constant and $\omega_M=\frac{1}{4}\omega_M^{AB}\Gamma_{A}\Gamma_{B}$ with $\omega_M^{AB}=\pt_Qe^{[A}e^{B]Q}-{\pt_Me^{[A}}_Qe^{B]Q}-e^{P[A}e^{Q|B]}{e^C}_M\pt_Pe_{CQ}$ is the spin connection. 
By varying the action with respect to the Dirac field $\bar\Psi$,  the equation of motion is written explicitly as
 \beq
\lt[ \gamma^\mu\pt_\mu+\gamma^5\lt(\pt_z+2\pt_z A \rt )-\eta \phi e^A \rt]\Psi	=0.
 \eeq
 
In order to decompose the 5-dimensional spinor into  4-dimensional parts and extra-dimensional parts, we employ a  chiral KK decomposition
$\Psi(x,z)=e^{-2A}\sum_n [\psi_{L,n}(x)f_{L,n}(z)+\psi_{R,n}(x)f_{R,n}(z)]$,
where $\psi_{L,n}=-\gamma^5\psi_{L,n}$ and $\psi_{R,n}=+\gamma^5\psi_{R,n}$ are the left-handed and right-handed components of a 4-dimensional Dirac field respectively. Then the 4-dimensional part $\psi_{L/R,n}$ satisfies a 4-dimensional Dirac equation $\gamma^\mu\pt_\mu\psi_{L/R,n}=m_n\psi_{R/L,n}$, and the extra-dimensional parts satisfy 
\begin{subequations}
\beqn
\lt(\pt_z+\mathcal{F}(z)\rt) f_{L,n}(z)&=&m_n f_{R,n}(z),\\
\lt(-\pt_z+\mathcal{F}(z)\rt) f_{R,n}(z)&=&m_n f_{L,n}(z).
\eeqn
\end{subequations}
where $\mathcal{F}(z)=\eta e^A \phi$. By assembling these two mixed first order equations, one arrives at two Schr\"odinger-like equations
\begin{subequations}\label{Fermion_SE}
\beqn
\lt(-\pt_z^2+V_{L}(z)\rt)f_{L,n}(z)&=&m_n^2f_{L,n}(z),\\
\lt(-\pt_z^2+V_{R}(z)\rt)f_{R,n}(z)&=&m_n^2f_{R,n}(z).
\eeqn
\end{subequations}
where the effective potentials are given by
$V_{L}(z)=\mathcal{F}^2(z)-\pt_z\mathcal{F}(z)$ and 
$V_{R}(z)=\mathcal{F}^2(z)+\pt_z\mathcal{F}(z)$.
Here the Hamiltonians of the Schr\"odinger equations (\ref{Fermion_SE}) can also be rewritten into the form of supersymmetric quantum mechanics,  namely, $H_L(z)=\bar\mathcal{ K}\bar\mathcal{ K}^\dagger$ and $H_R(z)=\bar\mathcal{ K}^\dagger\bar\mathcal{ K}$, with $\bar \mathcal{K}=-\pt_z+\mathcal{F}(z)$ and $\bar \mathcal{K}^\dagger=\pt_z+\mathcal{F}(z)$. So all the eigenvalues are nonnegative, and there is no tachyonic fermion state. Furthermore, supersymmetric quantum mechanics states that, except for the ground state, the eigenspectra of these two Hamiltonians are exactly the same, and it ensures that the 4-dimensional massive KK Dirac fermions can be reassembled by these left-handed and right-handed spinors.  The ground states are just the massless modes, which are easily got by setting $m_0=0$ in the Schr\"odinger equations,  and they read 
\begin{subequations}
\beqn
f_{L,0}&=& N_{L,0}e^{-\int \mathcal{F}(z)dz}=N_{L,0} e^{-\eta \int \phi(y) dy},\\
f_{R,0}&=& N_{R,0} e^{\int \mathcal{F}(z) dz}=N_{R,0}e^{\eta \int \phi(y) dy}.
\eeqn
\end{subequations}

 Only when a mode satisfies the normalization condition $\int_{-\infty}^{+\infty} e^{-A} f_{L/R,n}^2(y)dy =1$, will it be localized on the brane. With the brane solution (\ref{Brane_Solution}), we have  $e^{-A} f^2_{L,0}(y\rightarrow\pm\infty)\propto e^{\lt(\frac{k}{3\sqrt3}\mp\eta\rt) y}$ and $e^{-A} f^2_{R,0}(y\rightarrow\pm\infty)\propto e^{\lt(\frac{k}{3\sqrt3}\pm\eta\rt) y}$. Thus, for $\eta>\frac{k}{3\sqrt3}$, the left chiral zero mode is localizable, and for $\eta<-\frac{k}{3\sqrt3}$, the right one is localizable. This means that only one massless chiral mode can be localized on the brane, depending on the value of coupling constant $\eta$. It is well known that the left- and right-handed fermions transform differently under the electroweak gauge group in the Standard Model, thus via introducing 5-dimensional Dirac spinors $\Psi_i$ in different representations, and then setting the  coupling constants $\eta_i$ to choose the corresponding massless left- and right-handed Weyl spinors, a chiral gauge theory could be recovered on the brane \cite{Gherghetta2010}. 

\begin{figure*}[htb]
\begin{center}
\subfigure[$V_{L}(z)$ ]  {\label{Fig_Fermion_Potential_L}
\includegraphics[width=7cm,height=5cm]{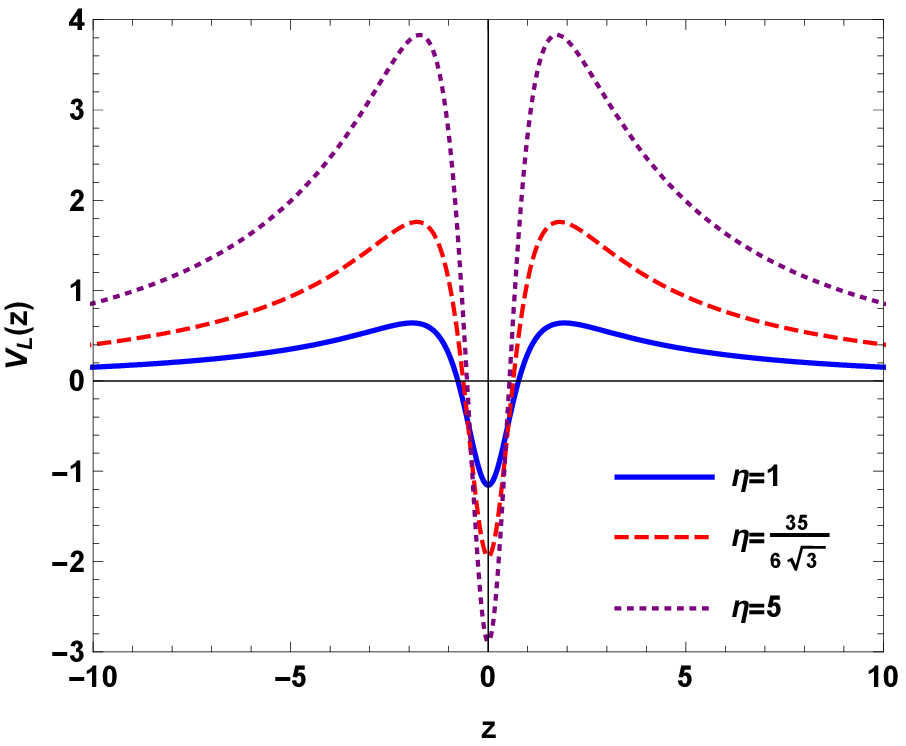}}
\qquad
\subfigure[$V_{R}(z)$]  {\label{Fermion_Potential_R}
\includegraphics[width=7cm,height=5cm]{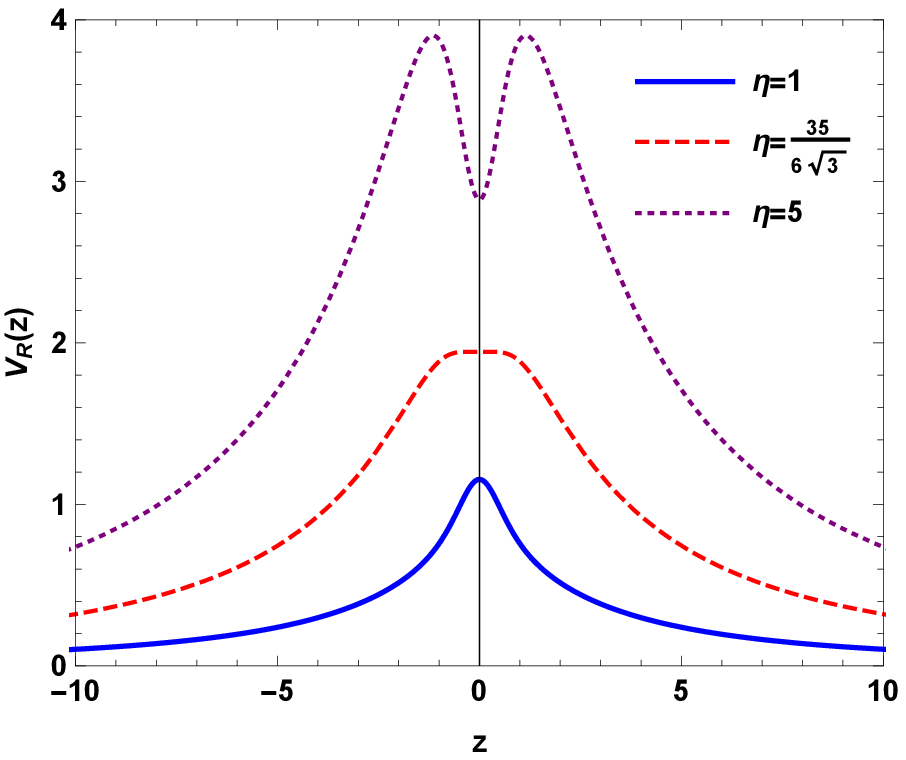}}
\end{center}
\caption{Plots of the Schr\"odinger potential $V_{L}(z)$ (a) and $V_{R}(z)$ (b). The parameter $k$ is set to $1$.}
\label{Fig_Fermion_Potential}
\end{figure*} 

As shown in Fig. \ref{Fig_Fermion_Potential}, the effective potentials $V_L$ and $V_R$ vanish at the boundaries, so only the massless mode is bound state and all the massive modes can escape into the extra dimension. For $-\frac{35k}{6\sqrt3}\le\eta\le\frac{35k}{6\sqrt3}$, there is no KK resonant state, since only $V_{L}$ or $V_R$ is volcano-like. However, for $\eta>\frac{35k}{6\sqrt3}$ (or $\eta<-\frac{35k}{6\sqrt3}$),  both $V_L$ and $V_R$ are volcano-like, so there would be some resonant (quasi-localized) states, which could  stay a longer time on the brane. In order to find these fermion resonances numerically, one can define a relative probability \cite{Liu2009c} : 
$P=\lt. \big(\int^{+z_b}_{-z_b}|f_{L/R,n}(z)|^2dz\big)\rt/\big(\int^{+10z_b}_{-10z_b}|f_{L/R,n}(z)|^2dz\big)$,
where $z_b$ is chosen as the coordinate referring to the maximum of the effective potential $V_{L/R}$. Since the effective potential is even, the wave function has either even or odd parity. Thus one can further impose the boundary conditions $f_{L/R,n}(0)=1, f'_{L/R,n}(0)=0$ for even parity  modes and $f_{L/R,n}(0)=0, f'_{L/R,n}(0)=1$ for odd, in order to solve the  Schr\"odinger equations (\ref{Fermion_SE}) numerically.

As an example, we numerically solve the KK resonances for $k=1$ and $\eta=4, 8, 10$, and the results are listed in Tab. \ref{Tab_Resonances}, where the lifetime $\tau$ of a KK resonance is estimated by $\tau\sim 1/\Gamma$, with $\Gamma = \delta m$ being the width of the half height of the resonant peak.
Since $\eta>\frac{35k}{6\sqrt3}$,  there is a localized massless left-handed ground state, whose parity is even, so the first left-handed KK resonance is parity-odd and first right-handed resonance parity-even.  For $\eta=4$, there is only one left-handed resonant state and one right-handed. For $\eta=8$,  there are two of each chiral resonances. For $\eta=10$,  the number increases to three. Therefore, as the Yukawa coupling becomes larger, more KK resonant states appear. This is due to the fact that the depth of potential well increases with the Yukawa coupling $\eta$, as shown in Fig. \ref{Fig_Fermion_Potential}. Furthermore, as the number of KK resonances increases, the relative probabilities of lower resonances become larger. Thus via enhancing the Yukawa interaction, more and more KK resonant fermions can be ``glued" on the brane and stay longer and longer.      

   \begin{table*}[!htb]
    \begin{center}
    \begin{tabular}{|c|c|c|c|c|c|c|}
     \hline
     $\eta$  & Chirality & Parity   & $m_{n}$  & $P$  & $\Gamma$                    & $\tau$   \\

      \hline
    
  \multirow{2}{*}{4}   &  $L$  & odd    & 1.651  
        &    0.299    &  0.310   &  3.224    \\ \cline{2-7}

        &  $R$   & even  & 1.643     
        & 0.385    &  0.254    & 3.933          \\    \hline 
          
     \multirow{4}{*}{8}   &       \multirow{2}{*}{$L$ }       & odd   & 2.690   
       & 0.963   &  $0.011$  &  91.196      \\  \cline{3-7}
       &             & even  & 3.254     & 0.290   
        & 0.213    & 4.692          \\  \cline{2-7}
             
          &   \multirow{2}{*}{$R$ }   & even  & 2.690  
             & 0.977    & 0.011   &  91.818   \\  \cline{3-7}
       &   & odd  & 3.243   
         & 0.339    & 0.213    &  4.692          \\ \hline 
   
        \multirow{6}{*}{10}   &       \multirow{3}{*}{$L$ }     & odd   & 3.094  
       & 0.997   & $8.887\times 10^{-4}$  & 1125.27 \\  \cline{3-7}
        &             & even  & 3.865     & 0.527   
        & 0.119    & 8.413           \\  \cline{3-7}
               &             & odd  & 4.259     & 0.183   
        & 0.293    & 3.416         \\  \cline{2-7}
             
          &   \multirow{3}{*}{$R$ }   & even  & 3.095  
             & 0.999    & $9.048\times 10^{-4}$    & 1105.17   \\  \cline{3-7}
      &   & odd  & 3.863   
         & 0.571    & 0.116    &  8.602        \\  \cline{3-7}
                &   & even  & 4.200   
         & 0.218    & 0.359    &  2.786         \\  \cline{3-7}      
      \hline
\end{tabular}\\
    \caption{The mass spectra $m_n$, relative probability $P$, width $\Gamma$, and lifetime $\tau$ of  the left- and right-handed KK resonances. The parameter $k$ is set to $1$.}
    \label{Tab_Resonances}
    \end{center}
    \end{table*}

\section{Conclusions}

In this work, we investigated a domain wall braneworld in the reduced Born-Infeld-$f(T)$ theory. By utilizing the first-order formalism, an analytic domain wall solution was obtained. The scalar field is always a single-kink regardless of the spacetime torsion, so there is no brane splitting in our case. This is different from the models built in the theory of $f(T)=T+k T^n$, where the branes can split for certain parameter region \cite{Yang2012b,Menezes2014}. Further, we considered the linear TT tensor perturbation against the background solution. It was found that the system is stable under the tensor perturbation and there is one localized massless graviton and continuous unlocalized massive gravitons. The localized massless mode recovers teleparallel gravity as the infrared theory on the brane. The massive gravitons decouple from the brane, so they contribute to a tiny correction with the behavior ${1}/{(k r)^{3}}$ to the Newtonian potential. At last, we investigated the (quasi-)localization of spin-1/2 fermion which couples to the background kink scalar via a Yukawa interaction. Only one of the two massless chiral modes can be localized on the brane depending on the value of the Yukawa coupling constant $\eta$. For the case of a volcano-like potential, all the massive states would escape into the extra dimension. However, via enhancing the Yukawa interaction, there are more KK fermion resonant states, and they can stay longer on the brane. Besides the TT tensor mode, there are (pseudo-)scalar and transverse (pseudo-)vector modes, and the model stability under the linear scalar and vector perturbations is left for our further investigation.

\section*{Acknowledgements}

The authors would like to thank Y.-P. Zhang for helpful discussions. This work was supported by the National Natural Science Foundation of China under Grant No. 11522541, No. 11375075 and No. 11747021.
K. Yang acknowledges the support of ``Fundamental Research Funds for the Central Universities" under Grant No. SWU-116052. Y.-X. Liu acknowledges the support of ``Fundamental Research Funds for the Central Universities" under Grant No. lzujbky-2016-k04.


\end{document}